\documentclass[12pt,a4paper]{article}
\usepackage{amsmath}
\usepackage{graphicx}
\usepackage{times}
\usepackage{cite}
\usepackage{ulem}
\usepackage{xcolor}
\usepackage{physics}
\begin{document}
\begin{titlepage}
\title{Cumulative activity of inelastic events  under hadron collisions}
\author{ S.M. Troshin, N.E. Tyurin\\[1ex]
\small  \it NRC ``Kurchatov Institute''--IHEP\\
\small  \it Protvino, 142281, Russian Federation,\\
\small Sergey.Troshin@ihep.ru
}
\normalsize
\date{}
\maketitle

\begin{abstract}
  We introduce the notion of  cumulative activity for inelastic events generated under hadron collisions, discuss its energy dependence and connection with the reflective scattering mode.  These issues
   are relevant for enlightening the asymptotic   dynamics in view of the LHC  measurements.
\end{abstract}
\end{titlepage}
\setcounter{page}{2}
\section{Introduction}
Hadrons are  composite, extended objects and their formfactors are  described by nontrivial functions. A seemingly natural expectation  is  an  increase  of the relative weight  of the inelastic interactions with the  collision energy. However,  the experimental results do not support  these naive  expectations.  A significantly increasing relative contribution of the elastic scattering events  to $pp$--interactions was observed, i.e. the ratio of  elastic to total cross-sections $\sigma_{el}(s)/\sigma_{tot}(s)$  inreases  with  energy, while the ratio $\sigma_{inel}(s)/\sigma_{tot}(s)$ decreases respectively. Those trends  have been  confirmed by the recent measurements at $\sqrt{s}=13$ TeV \cite {rat13}.  The divergence between experiment and the theoretical expectation is not surprising because of neglect the  color confinement effects under the use of hadron generation mechanism accounting the new channels.  This highlights the role of confinement  as the energy of hadron collisions  increases.

The hadron scattering amplitude as well as  the elastic and inelastic overlap functions reveal both the hadron geometry contribution and  dynamics of the constituents 
 under  hadron collision. 
The LHC  experiments have led to  discovery of collective effects  in  small systems, e.g. in 
$pp$-collisions with high multiplicities events \cite{weili}. To reveal the dynamics of a collective state formation one needs  to determine the impact parameter values associated with the particular events. This assumes reconstruction of the respective impact parameter values from the experimental observables since the impact parameter is  not directly measurable quantity. The notion of centrality   becomes an important issue when studing the small systems. Indeed, the centrality determines degree of the collision peripherality and is to be related to the collision geometry.

The centrality definition in the case of hadron interactions   \cite{cent}: 
\begin{equation}\label{cent}
	c^b(s)\equiv\frac{\sigma^b_{tot}(s)}{\sigma_{tot}(s)},
\end{equation} 
where 
\begin{equation}
\sigma^b_{tot}(s)=8\pi\int_0^b \mbox{Im} f(s,b')b'db',	
\end{equation}
$\sigma^b_{tot}(s)\to \sigma_{tot}(s)$ at $b\to\infty$ and $f(s,b)$ is elastic scattering amplitude in the impact parameter representation.
Eq. (\ref{cent})  takes into account 
presumed decoupling of   elastic scattering  from  multiparticle production processes.
This decoupling starts to occur  at small values of the impact parameter $b$ first and then expands to larger values of $b$ while collision energy increases. Such a behavior corresponds to increasing self--dumping of  inelastic contributions to the unitarity equation \cite{baker}. The knowledge of the decoupling dynamics   is essential   for development of QCD in the nonperturbative sector.

The above definition of centrality for the case of hadron scattering   has been given on base of the paper \cite{rog}. The important issues under its extension to the case of hadron scattering
are the  shape, energy density and color charge density fluctuations in the proton and their energy evolution \cite{alba,hei}.
Leaving aside those interesting but difficult and model--dependent issues, we  just assume that definition for centrality, Eq. (\ref{cent}),  is working for the case of hadron scattering and it allows one to reconstruct the respective impact parameter value. 

We duscuss here the cumulative properties of the inelastic events. To do this we introduce in the following the $b$--dependent quantity $a_{inel}^b(s)$, Eq. (\ref{act}), hereinafter reffered to as the cumulative inelastic activity.

Contrary to centrality,  the cumulative inelastic activity  obtains an essential contribution from the collision dynamics. The function $a_{inel}^b(s)$ being a way to express degree of peripherality for the inelastic events is closely related to the inelastic overlap function . The reflective scattering mode \cite{refl} appears in the developing peripheral impact parameter profile of $h_{inel}(s,b)$ at high energies, and  decreasing energy dependence of $a_{inel}^b(s)$ is an essential indication of  the reflective scattering mode presence \cite{jpg22}.

Thus, both the centrality and cumulative   inelastic activity  correspond to the different aspects of hadron interactions and they are complimentary quantities. Centrality  essentially reflects   geometry of the collision while the cumulative inelastic activity  is related to dynamics of the inelastic hadron interactions. 
\section{Energy dependence of cumulative inelastic activity  in hadron interactions}
Definition of cumulative inelastic activity in hadron interactions  goes back to the definition of centrality under    nuclei interactions (see e.g. \cite{rog}):
\begin{equation}\label{act}
a^b_{inel}(s)\equiv\frac{\sigma^b_{inel}(s)}{\sigma_{inel}(s)},
\end{equation}
where 
\textcolor{blue}{
\begin{equation}
\sigma^b_{inel}(s)=2\pi \int_0^b\sigma_{inel}(s,b')b'db'
\end{equation}
and  $\sigma_{inel}(s,b)$ is the differential contribution to the inelastic cross--section collisions  from the impact parameter $\vb{b}$  ($b\equiv|\vb{b}|$). 
The dimensionless function $\sigma_{inel}(s,b)$ ($\sigma_{inel}(s,b)\equiv d\sigma_{inel}/d^2\vb{b}$) is proportianal to the inelastic overlap function $h_{inel}(s,b)$, namely $$\sigma_{inel}(s,b)=4h_{inel}(s,b),$$} where 
\begin{equation}\label{un}
h_{inel}(s,b)=\mbox{Im}f(s,b)-|f(s,b)|^2.	
\end{equation}
and  $f(s,b)$ is the elastic scattering amplitude. 
The function $a^b_{inel}(s)$ is positive, it can variate from zero to unity and:
\begin{equation}\label{pin}
\frac{\partial a^b_{inel}(s)}{\partial b}=\frac{8\pi b}{\sigma_{inel}(s)} h_{inel}(s,b).
\end{equation}
Thus, cumulative inelastic activity $a^b_{inel}(s)$ is, by definition, a nonnegative and is limited by unity from above. Evidently that
$a^b_{inel}(s)\to 1$ when $b\to\infty$.

 The function $a^b_{inel}(s)$ accumulates contributions  of the inelastic events  generated under hadron collisions with  impact parameters in the interval $[0,b]$. It combines the effects of collision geometry with dynamical effects emphasizing the role  of hadron production mechanism and its energy dependence.    The differential inelastic activity is determined by the inelastic overlap function  $h_{inel}(s,b)$, Eq. (\ref{pin}).
 
 The proposal to use a cumulative definition of the inelastic activity is related to  probabalistic nature of the impact parameter  reconstruction from the centrality which in its turn has also both the cumulative and geometrical nature. The combined study of centrality and cumulative inelastic activity in hadron interactions  seems to be a useful tool for the  dynamics study and the hadron production models discrimination. 
 \textcolor{blue}{
 Indeed, the inelastic differential contribution $\sigma_{inel}(s,b)$  can be calculated from the experimentally measured  cumulative inelastic
activity\footnote{The relevant measurements are related to the respective multiplicity of produced charged hadrons.} at fixed values of centrality (Eq. \ref{pin}).}

Unitarity equation can be rewritten in the following form \cite{web}:
\begin{equation}\label{dif}
\sigma_{inel}(s,b)=\sigma_{tot}(s,b)-\sigma_{el}(s,b)=\sum_{n=3}^\infty \sigma_{n}(s,b).
\end{equation}
Magnitude of  the cumulative inelastic activity correlates with charged hadron production multiplicity and implies a radial flow of this quantity with energy increase from  center  to  periphery of the  interaction region  when the reflective scattering mode appears. Such flow can be interpreted as a local flow of  entropy in the interaction region under energy increase. This flow is absent in the absorptive scattering mode and is a specific feature of the relective scattering mode.

 To demonstrate a significant role of hadron interaction dynamics, namely due to the reflective scattering mode, and to find out energy dependence of the cumulative inelastic  activity we use the unitarization scheme and representation of the scattering amplitude $f(s,b)$ in the rational form.  Such an approach allows  variation of the amplitude in the whole interval allowed by unitarity \cite{umat}.
 In this case the $S$--matrix element $S(s,b)$ is written as the Cayley transform mapping nonnegative real numbers (in the pure imaginary scattering amplitude $f\to if$) to the interval $ [-1, 1]$\footnote{This one-to-one transform maps upper 
 	half-- plane into a unit circle in case the both $U$ and $S$  are complex functions and the value $S=0$ is reached at finite values of the energy and impact parameter.} :
 \begin{equation}
 S(s,b)=\frac{1-U(s,b)}{1+U(s,b)}. \label{umi}
 \end{equation}
  The real, nonnegative function $U(s,b)$ is considered as an input or bare amplitude which is subject to the unitarization procedure.
 Most of  the models used to construct a particular functional form of $U(s,b)$ provide  monotonically increasing energy
 dependence of the function $U(s,b)$ (e.g. power-like one) and its exponential decrease with  the impact parameter (due to analyticity in the Lehmann-Martin ellipse).
 The energy value  corresponding to  complete absorption of the initial state
 at  central collisions  $S(s,b)|_{b=0}=0$
 is denoted as $s_r$ and  determined by the  equation
 $U(s_r,b)|_{b=0}=1$\footnote{The  numerical estimates  of $s_r$ have given the value  ${s_r^{1/2}}=2-3$ $TeV$ at the pre--LHC time \cite{srvalue}.}.
 At  $s\leq s_r$  the scattering in the whole range of impact parameter variation
 has a shadow nature and   higher multiplicities  are associated with a more central collisions in the geometrical models.

The rational expression for the function $S(s,b)$, Eq. (\ref{umi}),  is consistent with invariance of the inelastic overlap function $h_{inel}(s,b)$ under increase of the function $U(s,b)$ with energy.
 A wide class of the geometrical models allows one to assume that $U(s,b)$ has a factorized form (cf. \cite{factor} and references therein):
 \begin{equation}\label{usb}
 U(s,b)=g(s)\omega(b),
 \end{equation}
 where $g(s)\sim s^\lambda$ at large values of $s$, and this power dependence guarantees asymptotic growth of the total cross--section $\sigma_{tot}\sim \ln^2 s$. Such factorized form, Eq. (\ref{usb}),  corresponds to a common reason of the  energy growth of the total cross--sections and diffraction cone slope in the  elastic scattering. 
 
 The particular simple form of the function $\omega(b)\sim e^{-\mu b}$ has been chosen to meet the analytical properties of the scattering amplitude. This choise of  $\omega(b)$ is also assumed by the physical picture grounded on its
 representation as a convolution of the two energy--independent hadron pionic-type matter distributions: 
 \begin{equation}
 \omega (b)\sim D_1\otimes D_2\equiv \int d {\bf b}_1 D_1({\bf b}_1)D_2({\bf b}-{\bf b}_1).
 \end{equation}
  Parameter $\mu$ should be then equal  to the doubled value of  pion mass.
Figure 1 illustrates  the notion of centrality  in hadron scattering in transverse plane.
 \begin{figure}[hbt]
 	\vspace{-1cm}
 	\hspace{-1cm}
 	\resizebox{16cm}{!}{\includegraphics{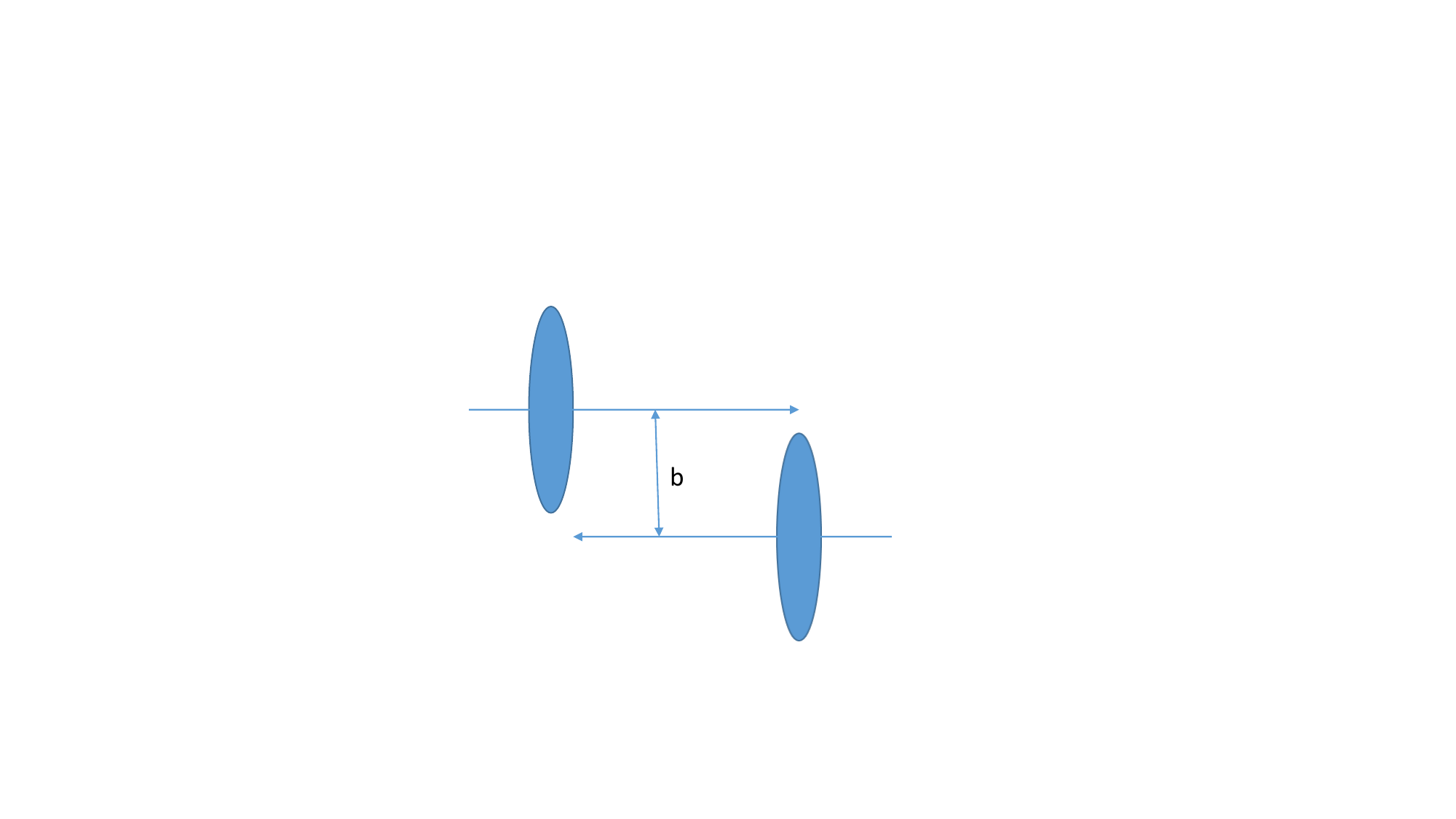}}		
 	\vspace{-2cm}
 	\caption{Schematic view of  hadron scattering  with the  impact parameter $b$ in the geometrical  models (cf. e. g.  \cite{heis, chy, low,  sof}) .}	
 \end{figure}	 
  \\
The explicit form of the inelastic overlap function   $$h_{inel}(s,b)=\frac{1}{4}(1-|S(s,b)|^2)$$ 
 when the  $U(s,b)$ is chosen in the form of Eq. (\ref{usb}) allows one to calculate  the cumulative inelastic activity given by Eq. (\ref{act}): 
 \begin{equation}
 \label{cbh}
 a^b_{inel}(s) =\frac{1}{\ln[1+g(s)]} \left[\ln\frac{1+g(s)}{1+g(s)e^{-\mu b}}-\mu b \frac{g(s)e^{-\mu b}}{1+g(s)e^{-\mu b}}\right].
 \end{equation}
 It leads to conclusion that at $s\to \infty$ and fixed value of $b$ one would expect strong energy decrease of the function $a^b_{inel}(s)$, i.e.
 \begin{equation}
 \label{asymp}
 a^b_{inel}(s)\sim \frac{1}{s^\lambda\ln(s)}.
 \end{equation}
It happens due to a power--like decrease of $\sigma^b_{inel}(s)$ and that of $\sigma_{inel}(s)\sim\ln s$. Power--like
decrease of $\sigma^b_{inel}(s)$ is correlated with the respective decrease of inelasticity observed in a series of cosmic ray experiments \cite{he}.  

Power--like decrease of the cumulative inelastic activity, Eq.(\ref{asymp}),  is associated with self--dumping of the inelastic channels and can be interpreted as a depletion of the respective radiating state.
Such power--like decrease is absent in the absorptive mode.  In the latter case,  the decrease is a logarithmic  at finite values of $b$ and the ratio of cumulative inelastic   activity and centrality $$r_{inel}^b(s)\equiv a^b_{inel}(s)/c^b(s)$$ does not vanish at $s\to\infty$ and $b$-fixed:
\begin{equation}\label{abs}
r_{inel}^b(s) \to 1.
 \end{equation}
It is a result of  the black disc limit saturation, i.e. it is  due to the same deacreasing logarithmic energy dependence of the both centrality and cumulative inelastic activty. 

The function $r_{inel}^b(s)$ is the ratio of the weighted cumulative quantities and characterizes relative cumulative contribution of the inelastic events for the given values of  $s$ and $b$.

The reflective scattering mode corresponds to vanishing   ratio $r_{inel}^b(s)$:
\begin{equation}\label{ref}
	r_{inel}^b(s) \to 0
\end{equation}
at $s\to\infty$.
We note that characteristic feature of centrality is its  weak energy dependence  in the both scattering modes. 

The impact parameter estimation problem for a given $pp$-collision event from the experimental data has been considered in \cite{cent}. It is  not a one--to--one reconstruction procedure. It is evident that the  event classification by multiplicity of the final state  is not appropriate for that purpose since a contribution of the elastic channel is then almost neglected. 
 The most relevant observable seems to be a sum of the transverse energies of the final state particles measured in a detector capable to register both elastic and inelastic events. Instead, the multiplicity measurements seem to be a tool for  estimation of  the cumulative activity of the  inelastic events.
 It seems that the CMS and TOTEM experiments \cite{cms, tot} are adopted  for   resolution of this problem.

 \section{Conclusion}
  
 The   features associated with energy and $b$--dependencies of the inelastic hadron interactions are considered in this note.
 Any quantity integrated over the {\it whole} region of its $b$--variation  (i.e. those taken at $-t=0$) is not sensitive to details of the  $b$-- and/or $t$--behaviour, respectively, and therefore cannot provide complete relevant information.  
 
 The $b$-dependent quantities are responsible for a number of the important  conclusions such as  peripheral form of the inelastic overlap function contrary to a central distribution of the elastic overlap function over the impact parameter. However,  the use of $b$-dependent quantities implies reduced statistical significance.
 It makes the results disputable. In that respect, the use of cumulative quantities allows one to resolve partly the case.
   The cumulative inelastic activity and  centrality extracted from the experimental data can be used for the elastic scattering amplitude reconstruction in the impact parameter space.  
   
   Studies of the ratio $r^b_{inel}(s)$  can serve as a tool for discrimination  of the particular forms of  asymptotics related to absorptive or reflective scattering modes. Decrease with energy of the ratio $r_{inel}^b(s)$  is to be considered in favor of the reflective mode presence. Such a behavior arises due to particular unitarization mechanism.

   The unitarization as a way to satisfy unitarity constraints  serves also to mitigation of strong energy dependence of the input under transition to the final scattering  amplitude. 
   Assumed strong  dependence at the level of the input reflects  the observed effects of  soft interactions dynamics while the unitarization procedure results in  additional conclusions.
   
   Asymptotic saturation of unitarity\footnote{Such saturation is provided, for example,  by Eqs. (\ref{umi}) and (\ref{usb}). Those relations guarantee  unitarity saturation and Eq. (\ref{ref})  is, in fact,  a result of this property. Saturation of unitarity corresponds to the principle of maximal strength of strong interactions proposed long ago by Chew and Frautchi \cite{chew}.}     leads to energy decrease of the ratio $r_{inel}^b(s)$   with corresponding evolution of hadron interaction region  towards black ring  picture. The  reflective ability  of the  inner region increases with energy towards unity \cite{mpla23} while the diameter and  ring width  grow up with energy  like $\ln s$. 
   
   Observation of the ridge in $pp$-interactions in the events with highgest multiplicities \cite{ridge} is in favor of the reflective scattering mode formation at the LHC energies. Indeed, such events correspond to the maximal absorption which moves into the region of nonzero impact paramers when energy increases, i.e. to peripheral collisions. It results in rotation of the deconfined matter in the hadron interaction region, and this rotation  manifests itself  as the ridge in the two-particle correlation function. Thus, the {\it emerging} phenomenon of ridge  formation (long--range two--particle correlations) discovered at the LHC in $pp$--interactions (see \cite{weili} for brief review of large variety of theoretical models) can be associated with  the reflecting scattering mode appearance. This mode occurs due to implementation of unitarity saturation and related production of a maximal number of the secondary particles at nonzero values of the impact parameter. Importance of collision peripherality for the two--particle long--range correlations appearance is confirmed by the ridge observation in the {\it peripheral} nuclear collisions (centrality controlled) at RHIC and LHC \cite{cms}.
 
Further search for  experimental manifestations of the reflective  mode in elastic and inelastic hadron interactions at available energies is   an important issue.  The centrality measurements combined with  measurements of the cumulative inelastic activity in $pp$-collisions  represent a promising way for achiving that purpose.  

The reflective  scattering mode would be associated with observation of a strongly decreasing cumulative inelastic activity with energy at fixed centrality values. Decrease of the cumulative inelastic activity at small impact parameters (proportional to $h_{inel}(s,0)$ ) already received the experimental justification at $\sqrt{s}=13$ TeV  \cite{jpg22}.  However,   a firm experimental  confirmation and further experimental sudies are needed  at the LHC energies and beyond. Since the cumulative inelastic activity is expected to start its decrease with energy at $b$ values close to zero, the  studies should be first limited to  small values  (0-5\%) of centrality.
 
 Thus, experimental studies of  the energy dependence of multiplicity under fixed value of centrality measured in its turn through  summing  up the transverse energies of the final state particles would be  rather promising for  discrimination between these scattering modes and further study of the    unitarization role.

\end{document}